\documentclass{PoS}

\PoS{PoS(HEP2005)336}

\title{Is FLARE for Solar flare?}

\ShortTitle{Is FLARE for flare?}

\author{\speaker{Daniele Fargion}\\
       Phys.Depart.Rome Univ.1,
      and INFN,Pl.A.Moro 2\\
       E-mail: \email{daniele.fargion@roma1.infn.it}}

\dedicated{to the Memory of  Sgt. Jonathan Evron, sacrificed on 2
November 2005}

 \abstract{The Fermi Lab Liquid ARgon experiment, FLARE, a
huge neutrino argon-liquid  project detector of $50$ kt mass,
might in a near future enlarge the neutrino telescope accuracy
revealing in detail solar, supernova, atmospheric as well as
largest solar flares neutrino. Indeed the solar energetic ($ E_p
> 100 MeVs $) flare particles (protons, $\alpha$)  while
scattering among themselves or hitting the solar atmosphere must
produce on sun prompt charged pions, whose decay (as well as their
sequent muon decays) into secondaries is source of  a copious
solar neutrino "flare" (at tens or hundreds MeV energy).  These
brief (minutes) neutrino "burst" at largest flare peak may
overcome by three to five order of magnitude the steady
atmospheric neutrino noise on the Earth, possibly leading to their
emergence and detection above the thresholds.  The largest prompt
"burst" solar neutrino flare may be detected in future FLARE
neutrino  $\nu$ detectors both in electron and positron and
possibly in its $muon$ pair neutrino component. Our estimate for
the recent and exceptional October - November $2003$ solar flares
and last January $20th$ 2005 exceptional flare might lead to a few
events for future FLARE or near unity for present
Super-KamiokandeII. The $\nu$ spectra may reflect the neutrino
flavor oscillations and mixing in flight. In neutrino detectors a
surprising (correlated) muon appearance may occur for a hard
(${E}_{{\nu}_{\mu}} \ge  200 MeV$) flare spectra, while a rarer
$\tau$ appearance may even marginally take place for
(${E}_{{\nu}_{\mu}} \rightarrow {E}_{{\nu}_{\tau}}\ge 4 GeV$)
spectra. A comparison of the solar neutrino flare signal with
other neutrino foreground is estimated: it offer the first
opportunity to a $muon$ neutrino astronomy, a much rarer $tau$
appearance, an independent road map to disentangle the neutrino
flavor puzzles, as well a prompt alarm system for  dangerous solar
flare eruptions. }

\FullConference{International Europhysics Conference on High Energy Physics\\
         July 21st - 27th 2005\\
         Lisboa, Portugal}

\begin{document}

\section{ FLARE signals for Up-Down Solar Neutrino Flare and  Flavors}

The recent peculiar solar flares on October-November $2003$ and
January $2005$ were source of high energetic charged particles :
large fraction of these {\itshape{primary}} particles, became a
source of both neutrons \cite{ref1} and {\itshape{secondary}}
kaons, $K^{\pm}$, pions, $\pi^{\pm}$ by their particle-particle
spallation on the Sun surface \cite{ref0}. Consequently,
$\mu^{\pm}$, muonic and electronic neutrinos and anti-neutrinos,
${\nu}_{\mu}$, $\bar{\nu}_{\mu}$, ${\nu}_{e}$, $\bar{\nu}_{e}$,
$\gamma$ rays, are released by the chain reactions $\pi^{\pm}
\rightarrow \mu^{\pm}+\nu_{\mu}(\bar{\nu}_{\mu})$, $\pi^{0}
\rightarrow 2\gamma$, $\mu^{\pm} \rightarrow
e^{\pm}+\nu_{e}(\bar{\nu}_{e})+ \nu_{\mu}(\bar{\nu}_{\mu})$
occurring on the sun atmosphere.
\begin{figure}[ht]
\input epsf
\includegraphics[width=15cm,height=4.1cm]{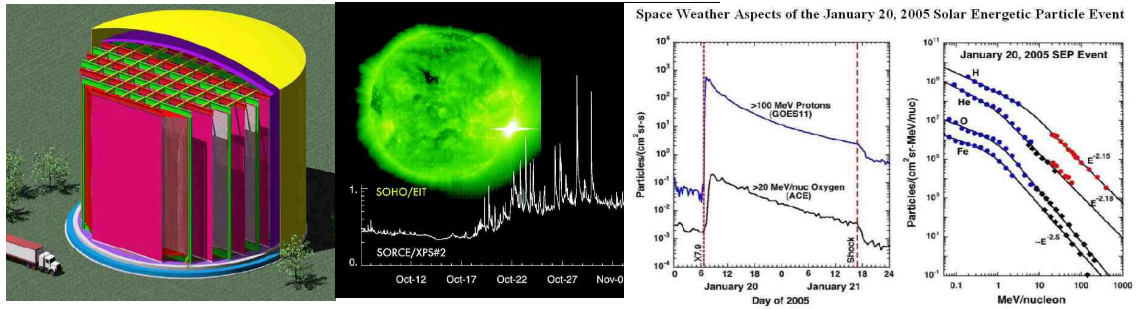}
\caption{The FLARE detector,  a real X-ray  on 4th Nov. 2003, a
solar flare spectra on 20 Jan. 2005}
\end{figure}
 There are two different sites for
these decays (see \cite{ref0}): A brief and sharp solar flare,
originated within the $solar$ corona and a diluted and delayed $
terrestrial$ neutrino flux, produced by flare particles hitting
the  Earth's  atmosphere; the additional solar flare particles,
hitting the terrestrial atmosphere increase the atmospheric
neutrino flux,  without any relevant consequence while later
cosmic (and atmospheric neutrino) flux should suffer of a
decrease,the Forbush effect, of little interest. The main and
first {\itshape{solar}} flare neutrinos reach the Earth with a
well defined directionality and within a narrow time range. The
corresponding average energies $<E_{{\nu}_{e}}>$,
$<E_{{\nu}_{\mu}}>$  are in principle a bit larger compared to an
event in the Earth's atmosphere since at low solar densities they
suffer negligible energy loss: $<E_{{\nu}_{e}}> $ $\simeq$ $50
MeV$, $<E_{{\nu}_{\mu}}> \simeq$ 100 $\div$ 200 MeV. The opposite
occur to downward flare. In the simplest approach, the main source
of pion production is $p+p\rightarrow {{\Delta}^{++}}n\rightarrow
p{{\pi}^{+}}n$;$p+p\rightarrow{{{\Delta}^{+}}p}^{\nearrow^{p+p+{{\pi}^{0}}}}_{\searrow_{p+n+{\pi}^{+}}}$
at the center of mass of the resonance ${\Delta}$ (whose mass
value is ${m}_{\Delta}=1232$ MeV). As an useful simplification
after the needed boost of the secondaries energies  one may assume
that the total pion $\pi^+ $ energy is equally distributed, in
average, in all its final remnants: ($\bar{\nu}_{\mu}$, ${e}^{+}$,
${\nu}_{e}$, ${\nu}_{\mu}$):${E}_{{\nu}_{\mu}} \geq
{E}_{{\bar{\nu}_{\mu}}} \simeq {E}_{{\nu}_{e}} \simeq
\frac{1}{4}{E}_{{\pi}^{+}}$. Similar nuclear reactions (at lower
probability) may also occur by proton-alfa scattering leading to:
$p+n\rightarrow {{\Delta}^{+}}n\rightarrow n{{\pi}^{+}}n$;
$p+n\rightarrow{{{\Delta}^{o}}p}^{\nearrow^{p+p+{{\pi}^{-}}}}_{\searrow_{p+n+{\pi}^{o}}}$.
Here we may neglect the ${\pi}^{-}$ additional role due to the
flavor mixing and the dominance of previous reactions ${\pi}^{+}$.
The flavor oscillation will lead to a decrease in the muon
component (as in atmospheric case) and it will make the electron
neutrino spectra a little harder. Indeed the oscillation length at
the energy considered is small respect Earth-Sun distance:
 $
 L_{\nu_{\mu}-\nu_{\tau}}=2.48 \cdot10^{9} \,cm \left(
 \frac{E_{\nu}}{10^{9}\,eV} \right) \left( \frac{\Delta m_{ij}^2
 }{(10^{-2} \,eV)^2} \right)^{-1} \ll D_{\oplus\odot}=1.5\cdot
 10^{13}cm$. For the  opposite reasons the tau appearance by
 atmospheric muon neutrino conversions is inhibited (at a few GeVs) on
 a much smaller Earth diameter distance. We take into account this flavor mixing by a conversion term
re-scaling the final muon neutrino signal and increasing the
electron spectra component \cite{ref0}. While at the birth place
the neutrino fluxes by positive charged pions $\pi^+$ are
$\Phi_{\nu_e}$:$\Phi_{\nu_{\mu}}$:$\Phi_{\nu_{\tau}}$ $= 1:1:0$,
after the mixing assuming a democratic number redistribution we
expect roughly
$\Phi_{\nu_e}$:$\Phi_{\nu_{\mu}}$:$\Phi_{\nu_{\tau}}$ $=
(\frac{2}{3}):(\frac{2}{3}):(\frac{2}{3})$. In a more detailed
balance the role of the most subtle  neutrino mixing $\Theta_{13}$
may be deforming the present averaged  flavor balance. On the
other side for the anti-neutrino fluxes we expect (neglecting
alfa-proton interactions) at the birth place:
$\Phi_{\overline{\nu_e}}$:$\Phi_{\overline{\nu_{\mu}}}$:$\Phi_{\overline{\nu_{\tau}}}$
$= 0:1:0$ while at their arrival (within a similar democratic
redistribution):$\Phi_{\overline{\nu_e}}$:$\Phi_{\overline{\nu_{\mu}}}$:$\Phi_{\overline{\nu_{\tau}}}$
$= (\frac{1}{3}):(\frac{1}{3}):(\frac{1}{3})$.  Because in
${\pi}$-${\mu}$ decay the ${\mu}$ neutrinos secondary are twice
the electron ones, the anti-electron neutrino flare energy is
(\cite{ref0}) from  the birth place on Sun up to the flavor mixed
states on Earth :
%%%%%%%%%%%%%%%%%%%%%%%%%eq.10(vecchia eq.10)%%%%%%%%%%%%%%%%%%%%%%%%%%%%%%%%%%%%%%%%%%%5
$
 {E}_{\bar{\nu}_{e}FL} \simeq\frac{{E}_{{\nu}_{{\mu}} FL}}{2}
\simeq 2.6\cdot{10}^{28}\left( \frac{{E}_{FL}}{{10}^{31}~erg}
\right)~erg. $ We scale our estimate with the total solar flare
energy ${E}_{FL}$ \cite{ref0}.
%%%%%%%%%%%%%%%%%%%%%%%%%%%%%%%%%%%%%%%%%%%%%%%%%%%%%%%%%%%%%%%%%%%%%%%%%%%%
The corresponding neutrino flare energy and number fluxes at sea
level (for the poor up-going flare) are:
%%%%%%%%%%%%%%%%%%%%%eq.11(vecchia eq.11)%%%%%%%%%%%%%%%%%%%%%%%%%%%%%%%%%%%%%%%%%%%%%%%%
$ {\Phi}_{\bar{\nu}_{e}FL} \simeq 9.15 \left(
\frac{{E}_{FL}}{{10}^{31}~erg} \right)~erg~{cm}^{-2} ;$
%%%%%%%%%%%%%%%%%%%%%%%%%%%%%%%%%%%%%%%%%%%%%%%%%%%%%%%%%%%%%%%%%%%%%
%%%%%%%%%%%%%%%%%%%%%eq.12(vecchia eq.12)%%%%%%%%%%%%%%%%%%%%%%%%%%%%%%%%%%%%%%%%%
$ {N}_{{\bar{\nu}_{e}}} \simeq \frac{{N}_{{{\nu}_{e}}}}{2} \simeq
5.7\cdot{10}^{4}\left(\frac{{E}_{FL}}{{10}^{31}~erg} \right)
\left( \frac{<{E}_{\bar{\nu}_{e}}>}{100~MeV} \right)^{-1} cm^{-2}.
$
%%%%%%%%%%%%%%%%%%%%%%%%%%%%%%%%%%%%%%%%%%%%%%%%%%%%%%%%%%%%%%%%%%

This neutrino number flux, associated to solar particle ejections
in $100$ s. time duration is larger by two-three order of
magnitude over the atmospheric one \cite{ref7} (one-two a second
for square cm. area) as it may be clearly seen from figure Fig.1
above for Jan.20th spectra. The daily cosmic atmospheric flux
induce a few neutrino events in SK (and three times more in
FLARE); therefore $if$ there are enough p-p interactions on Sun as
on Earth atmosphere, an integral flux of a day c.r. may correspond
to a brief cluster of events. We may expect in S.K (or better in
FLARE) detectors a signal at  the threshold level near one event
(or above a few). Let us consider horizontal flares (see fig. 1)
similar to horizontal and upward $\tau$ neutrino induced
air-showers at Earth Crust (see \cite{Fargion 2002}, \cite{Fargion
2004}). The solar neutrino flare production is enhanced by a
higher solar gas density where the flare propagate. Moreover a
beamed X-flare may suggest a primary beamed pion shower whose thin
jet naturally increases the neutrino signal. High energetic
protons flying downward (or horizontally) to the Sun exhibits
 an interaction probability ${P}_{d}$ larger than the previous
one (${P}_{up}$).  We  foresee \cite{ref0}, in general, that the
flare energy relations are: $E_{{\pi}FL}\equiv{\eta}E_{FL}\leq
E_{FL}$; $ E_{\bar{\nu}_{e}FL} \simeq\frac{E_{{\nu_{\mu}FL}}}{2}
\simeq 9.4\cdot10^{30}\eta \left( \frac{E_{FL}}{10^{32}~erg}
\right)~erg $. We expect a solar flare spectrum with an exponent
equal or larger (i.e. softer) than the cosmic ray proton spectrum
one (see fig.1). Estimated averaged neutrino energy
$<E_{\nu}>$ below GeV for down-going rich flare is \cite{ref0}:\\
$<N_{\bar{\nu}_{e}}> \: \simeq 1.72\cdot{10}^{6}{\eta} \left(
\frac{<E_{\bar{\nu}_{e}}>}{100~MeV} \right)^{-1} \left(
\frac{E_{FL}}{{10}^{31}~erg} \right)~{cm}^{-2} $; $
<N_{\bar{\nu}_{\mu}}> \: \simeq 4.12\cdot10^{6}{\eta} \left(
\frac{<E_{\bar{\nu}_{\mu}}>}{100~MeV} \right)^{-1} \left(
\frac{E_{FL}}{10^{31}~erg} \right)~{cm}^{-2} .$ The solar flare
neutrino events due to these number fluxes (following known
$\nu$-nucleons cross-sections at these energies
\cite{bemporad},\cite{strumia},\cite{Bodek}) at Super-Kamiokande
II is: $ {N}_{ev} \simeq 7.5 \cdot \eta \left(
\frac{E_{FL}}{10^{31} \, erg} \right) $ (for more details and
explanations see \cite{ref0}). The expectation event numbers at
SK-II for {\em solar neutrino burst} assuming, before mixing, a
more pessimistic detector thresholds calibrated with the observed
Supernovae 1987A event fluxes
\cite{ref0} has been evaluated:\\
 $ {N_{ev}}_{\bar{\nu}_{e}} \simeq
0.63{\eta}(\frac{\bar{E}_{\bar{\nu}_{e}}}{35
~MeV})(\frac{E_{FL}}{10^{31}~erg});~\bar{E}_{\bar{\nu}_{e}}\leq
100 ~MeV $; $ {N_{ev}}_{\bar{\nu}_{e}} \simeq
1.58{\eta}(\frac{E_{FL}}{10^{31}~erg});~
\bar{E}_{\bar{\nu}_{e}}\geq100-1000 ~MeV $;
${{N}_{ev}}_{\bar{\nu}_{\mu}} \simeq
3.58{\eta}(\frac{{E}_{FL}}{{10}^{31}~erg});~
\bar{E}_{\bar{\nu}_{\mu}}\geq 200-1000 ~MeV $; where the unknown
efficiency factor is ${\eta}\leq1$. The neutrino events in
Super-Kamiokande may be also recorded as stimulated beta decay on
oxygen nuclei \cite{ref8},\cite{ref0}. Let us underline the  role
of Solar Neutrino  Flavor mixing: the $\mu$ and $\tau$ appearance;
indeed the oscillation and mixing guarantee the consequent tau
flavor rise and the anti neutrino  electron component hardening
respect to the one at its birth.
 This increase the neutrino electron component while it will reduce
 the corresponding muon component leading to : ${\frac{\eta_{\mu}}{\eta_{e}}\simeq \frac{1}{2}}$  and  to
      ${N}_{{ev}_{\bar{\nu}_{\mu}}}\simeq {N}_{{ev}_{\bar{\nu}_{e}}}
       \simeq 2 (\frac{<{E}_{\bar{\nu}_{\mu}}>}{200~MeV})(\frac{<{E}_{FL}>}{{10}^{31}~erg})$ ;
     ${N}_{{ev}_{{\nu}_{\mu}}}\simeq {N}_{{ev}_{{\nu}_{e}}} \simeq 4 (\frac{<{E}_{{\nu}_{\mu}}>}{200~MeV})(\frac{<{E}_{FL}>}{{10}^{31}~erg})
     $,   as well as a comparable, ${\nu}_{e}$, ${\nu}_{\mu}$, $\bar{\nu}_{e}$, $\bar{\nu}_{\mu}$
        energy fluency and  spectra.
        At energies above a few hundred MeV proton-proton scattering may release
        hundred MeV muon neutrino able to create first muon neutrino from the sun.
         Moreover  above the $\tau$ threshold energy ${E}_{{\nu}_{\mu}} \geq 3.46 $ GeV a
         surprising (but very rare) $\tau$ appearance may occur:
          (${E}_{{\nu}_{\mu}} \rightarrow {E}_{{\nu}_{\tau}}$$\simeq  4 GeV$) flare spectra.
 Any positive evidence for such electron, muon, tau
  events will mark a new road to Neutrino Astrophysics, to be complementary to lower
neutrino energy from Sun and Supernov\ae. New larger generations
of neutrino detectors as FLARE will be more sensitive to such less
powerful, but more frequent and energetic solar flares, than to
the rarest galactic and even extragalactic supernov\ae~ events (as
the one from Andromeda) \cite{Fargion 2002}.  The background due
to energetic atmospheric neutrinos at the Japanese detector was
nearly $5.8$ event a day corresponding to a rate ${\Gamma\simeq
6.7 10^{-5}} s^{-1}$. The probability to find by chance one
neutrino event within a $1-2$ minute solar flare ${\Delta}t \simeq
10^2 s$ in that interval is $P\simeq\Gamma \cdot{{\Delta}T}\simeq
6.7 \cdot10^{-3}$. For a Poisson distribution the probability to
find  contained  $n=1,2, 3, 4, 5$ events in a  narrow time window
might reach extremely small values: $
 {{P}_{n}}\cong e^{-P}\cdot \frac{P^{n}}{n!}\simeq \frac{P^{n}}{n!}=( 6.7
\cdot 10^{-3}, 2.25 \cdot 10^{-5},  5 \cdot 10^{-8}, 8.39 \cdot
10^{-11}, 1.1\cdot10^{-13}). $ Therefore the possible presence of
one or more high energetic (tens-hundred MeVs) positrons or
positive muons  (twice as much because privileged $\pi^+$
primary),  as well as negative electrons or muons in
Super-Kamiokande or FLARE at X-flare onset time, may be a defined
signature of the solar neutrino flare. FLARE while tracking muon
and its electron inside may point to the solar flare $\nu_{\mu}$
into $\mu$ astronomy or even to the $\tau$ appearance, for hard
(${E}_{{\nu}_{\mu}} \rightarrow {E}_{{\nu}_{\tau}}> 4 GeV$) flare
spectra, and it may surprise soon  us. An $individual$ or $global$
time correlation with past large solar flares with  SK past
activity) record might hide already signals. We conclude that the
 neutrino detectors  as FLARE, UNO, HK might be at
the same time ideal laboratories for solar neutrino flare and
flavor mixing, as well as the most rapid alert system monitoring
the huge solar flare and coronal mass ejection, dangerous for
orbiting satellites and possibly lethal for on flying astronauts.
The prompt solar neutrino flare alarm may provide the needed time
to shut down delicate satellite systems and to protect astronauts
into the inner  and saver space-shuttle shields.

\begin{figure}[ht]
\input epsf
 \includegraphics[width=15cm,height=5.1cm]{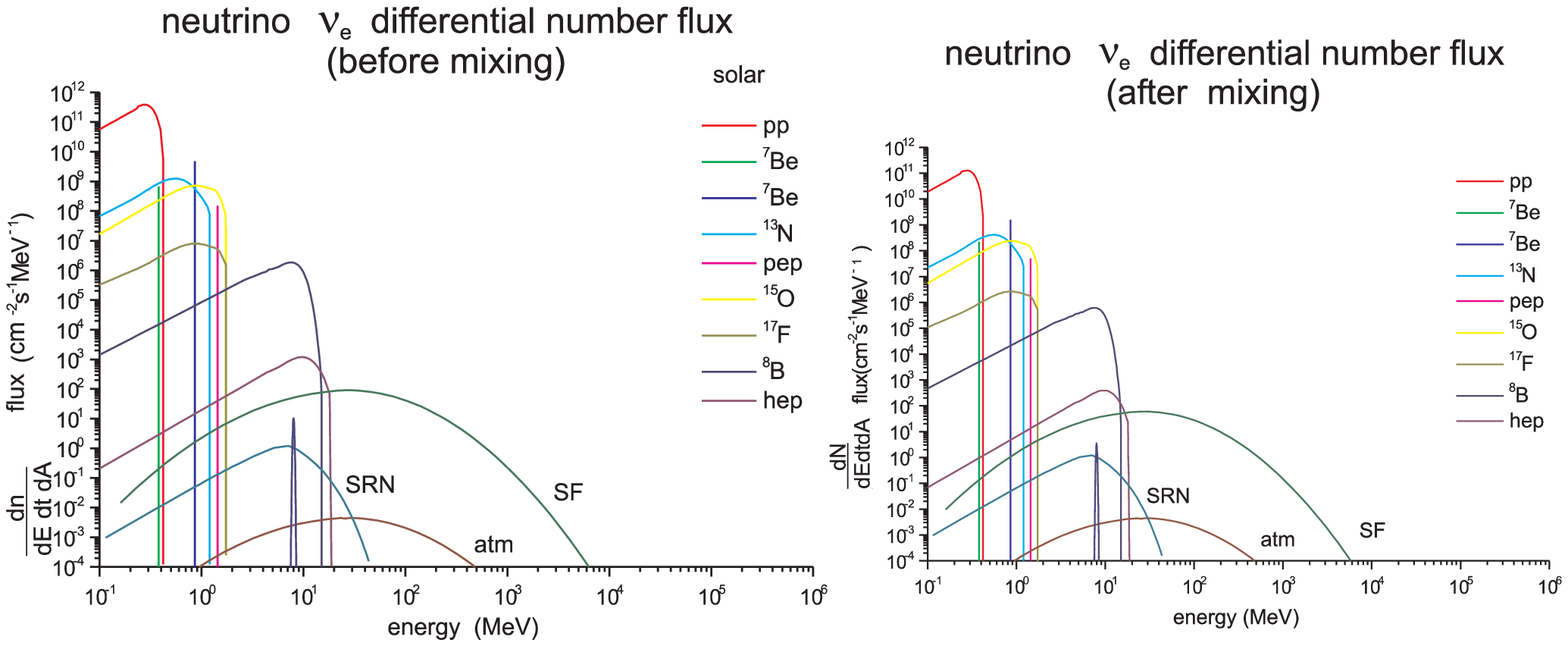}
\caption{Solar Neutrino $\nu_e$ flare before and after the flavor
mixing; note the thin $\nu_e$ peak due to prompt neutronization
flash of cosmic Supernovae, the solar neutrinos flux by known
nuclear activities as well as the expected Supernova Relic fluxes
and the atmospheric neutrino background.  The Solar flare at peak
activity hold nearly $100-1000$ $s$ and it has a  label  SF. Both
expectations  are described for an up-going (poor flux) or
down-going (richer flux) solar "burst" scenario. The primary solar
flare spectra is considered like the atmosphere one at least
within the energy windows $E_{{\nu}_{\mu}} \simeq 10^{-3} GeV$ up
to $ 10$ GeV as the 20 Jan. spectra shown above. }
\label{Fig:demo1}
\end{figure}

\begin{figure}[ht]
\input epsf
\includegraphics[width=15cm,height=5.1cm]{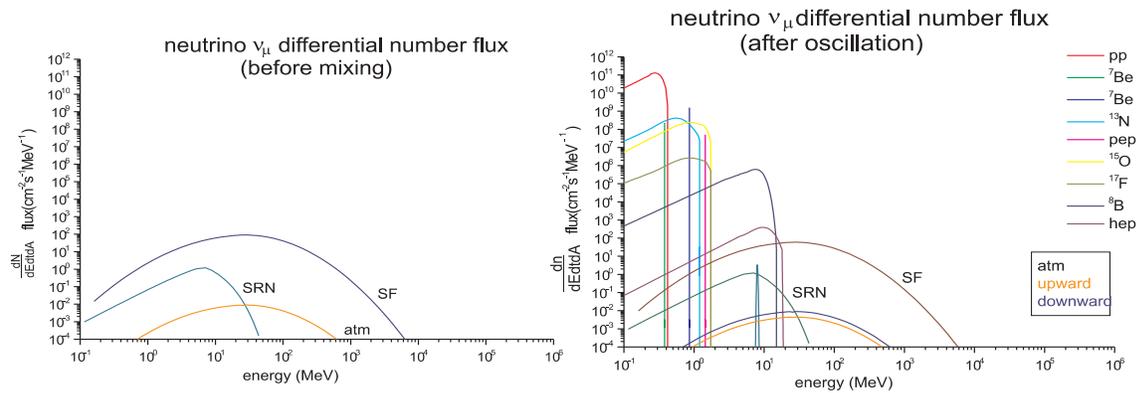}
\caption{As above Solar Neutrino $\nu_{\mu}$ before and after
mixing, as well as a comparable final $\nu_{\tau}$ flux; note the
presence of two different  fluxes (upward and down-ward
$\nu_{\mu}$ on Earth) due to the muon oscillation and
disappearance into $\tau$ flavor}
\end{figure}

   \begin{figure}[ht]
\input epsf
\begin{center}
\includegraphics[width=3cm,height=3.3cm]{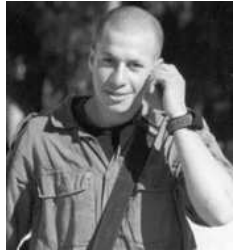}
\caption{The Sgt. Jonathan Evron, fallen  on 2nd November 2005
 only at 20 years age: his sunny and  flaring life, his humor and immense generosity  was inspiring the article,
devoted to his memory}
\end{center}
\end{figure}

\end{document}